\documentclass[amsmath,twocolumn,showpacs,showkeys]{revtex4}
\usepackage{amsmath,amsfonts,amssymb}

\def\a{\alpha}
\def\b{\beta}
\def\g{\gamma}
\def\vt{\vartheta}
\def\d{\delta}

\def\stareq{\stackrel{*}{=}}

\def\@{\partial_}

\def\chris#1#2#3{#1\brace #2 #3}
\def\negenspace{\kern-1.1em}

\def\sqr#1#2{{\vcenter{\hrule height.#2pt\hbox{\vrule width.#2pt
height#1pt \kern#1pt \vrule width.#2pt}\hrule height.#2pt}}}
\def\square{\mathchoice\sqr64\sqr64\sqr{4.2}3\sqr{3.0}3}

\usepackage{graphicx, epic, eepic, color}

\definecolor{wichtig}{rgb}{1,0,0} 
\definecolor{folge}{rgb}{0,0,1} 
\definecolor{liste}{rgb}{0,0.7,0} 

\definecolor{dark-green}{rgb}{0,0.7,0}
\definecolor{dark-blue}{rgb}{0,0.2,0.5}
\definecolor{med-blue}{rgb}{0,0.7,1}
\definecolor{mblue}{rgb}{0,0.2,1}
\definecolor{cnc}{rgb}{0.8,0,0}
\definecolor{light-red}{rgb}{1,0.8,0.8}
\definecolor{dark-yellow}{rgb}{1,0.8,0}
\definecolor{light-blue}{rgb}{0.8,0.9,1}
\definecolor{verylight-blue}{rgb}{0.93,0.95,1}
\definecolor{light-yellow}{rgb}{1,0.9,0.8}
\definecolor{grey}{gray}{0.88}

\def\bfx{\mathbf{x}}

\def\bfy{\mathbf{y}}

\def\bfz{\mathbf{z}}

\def\bfr{\mathbf{r}}

\def\bfk{\mathbf{k}}

\def\rhod{\rho_{\rm D}}

\begin{document}
\title{A formal framework for a nonlocal generalization of
  Einstein's theory of gravitation}

\author{Friedrich W. Hehl}
\email{hehl@thp.uni-koeln.de} 
\affiliation{Institute for Theoretical Physics, University of Cologne,
50923 K\"oln, Germany}
\affiliation{Department of Physics and Astronomy,
University of Missouri, Columbia, MO 65211, USA}
\author{Bahram Mashhoon}
\email{mashhoonb@missouri.edu}
\affiliation{Department of Physics and Astronomy,
University of Missouri, Columbia, MO 65211, USA}

\begin{abstract}

  The analogy between electrodynamics and the translational gauge
  theory of gravity is employed in this paper to develop an ansatz for
  a nonlocal generalization of Einstein's theory of gravitation.
  Working in the linear approximation, we show that the resulting
  nonlocal theory is equivalent to general relativity with ``dark
  matter''. The nature of the predicted ``dark matter'', which is the
  manifestation of the nonlocal character of gravity in our model, is
  briefly discussed. It is demonstrated that this approach can provide
  a basis for the Tohline-Kuhn treatment of the astrophysical evidence
  for dark matter.
\end{abstract}



\date{06 April 2009, {\it file NonlocalGrav40.tex}}

\maketitle

\section{Introduction}

In special relativity theory, the principle of locality has been the
subject of a detailed critical analysis and nonlocal correction terms
have been proposed that are induced by sufficiently high {\it
  accelerations} \cite{BahramAnnalen}. This nonlocal special
relativity has been applied to electrodynamics and the Dirac
equation. For a discussion of possible experimental tests of the
theory, we refer to \cite{BahramAnnalen} and the references cited
therein.

The next subject to be addressed from this nonlocal point of view is
the theory of gravitation. Einstein's theory of general relativity
(GR) is, by means of the equivalence principle, heuristically deduced
from special relativity. Therefore, if special relativity is
generalized to a nonlocal theory, then general relativity theory
cannot be exempt from this generalization process. But how should such
a generalization be implemented in Einstein's theory? Should one, for
example, require the connection to be a nonlocal expression in terms
of the metric? No obvious method suggests itself for a direct nonlocal
generalization of Einstein's theory. The underlying reason for this is
that the precise mathematical form of Einstein's principle of
equivalence in GR is {\it strictly} local; therefore, it cannot be
employed together with nonlocal special relativity to arrive at
nonlocal gravitation. On the other hand, various heuristic arguments
have been advanced in favor of a nonlocal classical theory of
gravitation (see, for instance, \cite{Bahram2007}).

The idea behind this paper exploits the analogy between {\it gauge
  theories.} On the one hand we know that electrodynamics, in the
framework of a gauged Dirac system, can be understood as a gauge
theory of the $U(1)$ group; see O'Raifeartaigh \cite{Lochlainn}.  On
the other hand it is known that a gauge theory of the {\it
  translation} group, for spinless matter, yields a teleparallelism
theory of gravity that, for a suitably chosen Lagrangian, is
equivalent to Einstein's theory; see, for instance, Nitsch et al.\
\cite{NitschErice,NitschHehl}. Consequently, if we cannot recognize a
direct method of generalizing Einstein's theory, it may be helpful to
start from this so-called teleparallel equivalent of general
relativity (GR$_{||}$) instead. This will indeed provide a way to
proceed to a nonlocal extension.

The analogy between these groups is pointed out in the rest of this
section and the resulting ansatz for nonlocal gravitation is explained
in Sec.\ II. To simplify matters, we work in the linear approximation
and work out the linearized nonlocal gravitational field equations in
Sec.\ III. In Sec.\ IV some subtleties are discussed in connection
with the use of variational principles in nonlocal field
theories. Section V is devoted to an interpretation of these nonlocal
equations in terms of the standard linearized general relativity
theory but in the presence of {\it ``dark matter''.} The properties of
the resulting ``dark matter'' are briefly pointed out. A discussion of
our results is contained in Sec.\ VI.  Various mathematical details
are relegated to the appendices. In this paper, spacetime indices run
from $0$ to $3$ and units are chosen such that $c = 1$. The Minkowski
metric tensor is given by {diag}$(1, -1, -1, -1)$. As indices, Latin
letters indicate holonomic coordinate indices, while Greek letters
indicate anholonomic Lorentz-frame (tetrad) indices.

\subsection{Electrodynamics}

The electromagnetic {\it excitation} ${\cal H}^{ij}=(D,H)=-{\cal
  H}^{ji}$, a contravariant tensor density, and the {\it field
  strength} $F_{ij}=(E,B)=-F_{ji}$, a covariant tensor, fulfill the
Maxwell equations
\begin{equation}\label{Max}
  \partial_j{\cal H}^{ij}={\cal J}^i\,,\qquad \partial_{[i}F_{jk]}=0\,,
\end{equation}
with ${\cal J}^i$ as electric current density. Here $i,j,...=0,1,2,3$
are (holonomic) coordinate indices. The homogeneous equations
(\ref{Max})$_2$ can be solved by the potential ansatz
\begin{equation}\label{pot}
F_{ij}=2\partial_{[i}A_{j]}.
\end{equation}
The two Maxwell equations have to be supplemented by a constitutive
law that relates the excitation to the field strength.  In
conventional vacuum electrodynamics, one takes the local, linear, and
isotropic `constitutive law'
\begin{equation}\label{vac} {\cal H}^{ij}=\left\langle
    \sqrt{-g}g^{k[i}g^{j]l}\right\rangle F_{kl}=\sqrt{-g}\,F^{ij}\,.
\end{equation}
This law is equally valid for the flat Minkowski spacetime of special
relativity as well as for the curved Riemannian spacetime of general
relativity. In the former case, in Cartesian coordinates, the metric
$g_{ij}$ is constant, in the latter case it becomes a field governed
by the Einstein equations. If we substitute Eq.~(\ref{vac}) into
Eq.~(\ref{Max})$_1$, then the Maxwell equations are expressed in terms
of the field strength $F_{ij}$ alone.

In a nonlocal theory, on the other hand, equation (\ref{vac}) is
generalized such that the nonlocal constitutive law expresses ${\cal
  H}^{ij}$ as a nonlocal spacetime relation in terms of $F_{kl}$. In
fact, it is possible to give an effective interpretation of Mashhoon's
nonlocal electrodynamics in this way \cite{Muench00}. That is, one
refers the excitation ${\cal H}^{\a\b} =e_i{}^\a e_j{}^\b {\cal
  H}^{ij}$, the field strength $F_{\a\b}=e^i{}_\a e^j{}_\b F_{ij}$,
and the metric $g_{\a\b}=e^i{}_\a e^j{}_\b\, g_{ij}$ to non-inertial
frames $e_\a=e^i{}_\a \@i$ that characterize the tetrads of
accelerated observers. With respect to such frames, the Maxwell
equations read \cite{Muench00,Birkbook}
\begin{eqnarray}\label{maxcomp}
 \partial_\beta { {\cal H}}{}^{\alpha\beta} -\frac 12  
  C_{\beta\gamma}{}^\alpha\, {{\cal H}}{}^{\gamma\beta}
+\frac 12  
  C_{\beta\gamma}{}^\g\, {{\cal H}}{}^{\a\beta}
&=&
  {{\cal J}}{}^{\alpha}\,,\\ \partial_{[\alpha}
  F_{\beta\gamma]}-C_{[\alpha\beta}{}^\delta F_{\gamma]\delta}&=&0\,.
\end{eqnarray}
Here $C_{\a\b}{}^\g:=2e^i{}_\a e^j{}_\b\@{[i}e_{j]}{}^\g$ is the
object of anholonomity (``structure constants'') measuring the
non-inertial nature of the frame. Thus in this approach to nonlocal
electrodynamics, the nonlocal expression that relates the excitation
to the field strength is given by \cite{Muench00}
\begin{eqnarray}\label{nonlocalMax}
{\cal H}^{\a\b}(x)&=&\sqrt{-g}g^{\a\g}g^{\b\d}F_{\g\d}(x)\nonumber\\ &&+
  \int\chi^{\a\b\g\d}(x,x')F_{\g\d}(x')d^4x'\,.
\end{eqnarray}
Eq.~(\ref{nonlocalMax}) represents a new axiom that, together with
the Maxwell equations (\ref{Max}), determines a nonlocal theory of
electrodynamics.

\subsection{Translational gauge theory of gravity}

Let us now turn to the gauge theory of the translation group, which
includes GR$_{||}$. We take the field equations from
\cite{HehlErice79,HehlHeld}; however, we change some conventions in
order to conform with \cite{PRs,Birkbook}. For the tensor calculus
background one should compare Schouten \cite{Schouten}. In Appendix~A
we provide a brief sketch of translational gauge theory formulated in
terms of the calculus of exterior differential forms. Recent works on
teleparallelism can be found, amongst others, in the articles of
Aldrovandi et al.\ \cite{Aldrovandi1}, Itin \cite{Itin1,Itin2}, Maluf
et al.\ \cite{Maluf1,Maluf2}, Mielke \cite{Egg}, Obukhov et al.\
\cite{YuriTele1,YuriTele2,YuriTele3}, So et al.\ \cite{So}, and Tung
et al.\ \cite{Tung}; moreover, the two monographs of Blagojevi\'c
\cite{Milutin} and Ort\'{\i}n \cite{Ortin} should be consulted as well
as the references cited therein. The inhomogeneous and the homogeneous
gravitational field equations are respectively given by
\begin{equation}\label{fieldeq}
  \partial_j{\cal H}^{ij}{}_\a-{\cal E}_\a{}^i\stackrel{*}{=}{\cal
    T}_\a{}^i\,,\qquad \partial_{[i} C_{jk]}{}^\a\stackrel{*}{=}0\,.
\end{equation}
The star over the equal sign means that the corresponding equation is
only valid in a suitable frame; this will be discussed in Sec.\ II.  The
nonlinear correction terms in Eq.~(\ref{fieldeq})$_1$ represent the
energy-momentum tensor density of the gravitational gauge field,
\begin{equation}\label{energy}
{\cal E}_\a{}^i:=-\frac 14  e^i{}_\a( C_{ jk}{}^\b
{\cal H}^{jk}{}_\b) + C_{\a k}{}^\b {\cal H}^{ik}{}_\b\,.
\end{equation}
As source, we have on the right-hand side of the inhomogeneous field
equation the energy-momentum tensor density of matter. In analogy with
Eq.~(\ref{pot}), the field strength $ C_{ij}{}^\a$ is defined in
terms of the potential $e_i{}^\a$ by
\begin{equation}\label{Omega}
 C_{ij}{}^\a=2\partial_{[i}e_{j]}{}^\a\,,
\end{equation}
so that the homogeneous gravitational field equations
(\ref{fieldeq})$_2$ are thus identically satisfied.

\begin{table} \label{Tab1}
  \renewcommand\arraystretch{1.3} $\begin{array}{|c|c|c|}\hline
    \text{Expression for}&\text{Electrodynamics}&\text{Transl.\
      Gauge Theory}\\ \hline \text{conservation law}&\@i{\cal J}^i=0&
    \@i{\cal T}_\a{}^i \sim0\\ \hline \text{inh.\ field eqs.}&\@j
    {\cal
      H}^{ij}={\cal J}^i&\@j{\cal H}^{ij}{}_\a\sim {\cal T}_\a{}^i\\
    \hline \text{force density}&{f}_i=F_{ij}{\cal J}^j&{ f}_i\sim
    C_{ij}{}^\a\,{\cal T}_\a{}^j\\ \hline \text{hom.\ field
      eqs.}&\@{[i}F_{jk]}=0&\@{[i} C_{jk]}{}^\a=0\\ \hline
    \text{potentials}&F_{ij}=2\@{[i}A_{j]}&
    C_{ij}{}^\a=2\@{[i}e_{j]}{}^\a \\ \hline
\end{array}$
\renewcommand\arraystretch{1}
\caption[]{Analogy between electrodynamics and gravitodynamics. The
  approximation sign employed in the last column indicates validity
  only in the {\it linear} regime.}
\end{table}

Clearly, there is a similarity between the electrodynamic and
gravitational cases, as indicated in Table I. In electrodynamics we
have one potential $A_i$ as covector, corresponding to the
one-parameter $U(1)$ group; in the gravitational case, because we have
four linearly independent translations in spacetime, we have four
covectors as potentials $e_i{}^\a$, the tetrad components with
$\a=\hat{0},\hat{1},\hat{2},\hat{3}$. Correspondingly, whereas we have
six components of the electromagnetic field strength $F_{ij}=-F_{ij}$,
we have $4\times 6$ components of the {\it gravitational field
  strength} $ C_{ij}{}^\a=- C_{ji}{}^\a$. If one compares
Eq.~(\ref{Max}) with Eq.~(\ref{fieldeq}), one recognizes the
quasi-Maxwellian nature of the gravitational case, only that we have
four times more field equations in Eq.~(\ref{fieldeq}) and that the
gravitational energy-momentum tensor density ${\cal E}_\a{}^i$ emerges
because all physical fields including gravity are gravitationally
`charged', that is, all fields carry energy-momentum.

The analog of Eq.~(\ref{vac}), the local, linear, and isotropic
constitutive law in GR$_{||}$ is \cite{NitschHehl,HehlHeld}
\begin{equation}\label{ten} {\cal
    H}^{ij}{}_\a\stackrel{*}{=}\frac{e}{\kappa}\left( \frac 12
    C^{ij}{}_\a - C_\a{}^{[ij]}+2 e^{[i}{}_\a C^{j]
      \gamma}{}_\gamma\right)\,,
\end{equation}
where $e:=\det(e_i{}^\a)$. Here $\kappa=8\pi G$, where $G$ is Newton's
constant of gravitation.  If Eq.~(\ref{ten}) is substituted into the
inhomogeneous field equations (\ref{fieldeq})$_1$, the resulting
equations have been shown to be equivalent to the Einstein equations
\cite{HehlHeld}. Taking the electromagnetic case (\ref{nonlocalMax})
as a prototype, we tentatively expect that a suitable nonlocal
generalization of the constitutive relation (\ref{ten}) would lead to
a nonlocal generalization of Einstein's theory of gravitation. It is
interesting to contemplate the nature of this analogy, which is the
basis of the present paper. The {\it linear} and possibly {\it
  nonlocal} electrodynamic constitutive law is generally valid for
sufficiently weak electromagnetic fields, since the constitutive
tensor is assumed to be independent of the field strength. Therefore,
we expect that the same holds in the linear approximation for the
nonlocal theory of gravitation that we develop below on the basis of
this analogy.

Finally, we must mention that this approach to nonlocal gravitation is
essentially different from other nonlocal modifications of general
relativity; see, for example, \cite{Soussa2003}, \cite{Barvinsky2003},
and the references cited therein.

\section{Ansatz for a nonlocal theory of gravity}

The arena for teleparallelism is the Weitzenb\"ock spacetime in which
we have the orthonormal tetrad (coframe) $\vt^\a=e_i{}^\a dx^i$. Then
the local (anholonomic) metric is
$\eta_{\a\b}=\text{diag}(1,-1,-1,-1)$ and we can determine the
(holonomic) coordinate components of the metric by
\begin{equation}\label{holmetric}
g_{ij}=\eta_{\a\b} e_i{}^\a e_j{}^\b\,.
\end{equation}
Simple algebra yields $e:=\text{det}(e_i{}^\a)=\sqrt{-g}$. 

If a coframe is a coordinate frame, then the object of anholonomity 
\begin{equation}\label{anhol}
  C^\a:=d\vt^\a
\end{equation}
vanishes; in general, $C^\a$ is a 2-form which decomposes according to
$C^\a=\frac 12 C_{ij}{}^\a dx^i\wedge dx^j$ such that Eq.~(\ref{anhol}) in
components results in Eq.~(\ref{Omega}).

The connection 1-form $\Gamma^{\a\b}=-\Gamma^{\b\a}=
\Gamma_i{}^{\a\b}dx^i$ in a Weitzenb\"ock spacetime is
teleparallel. That is, we can introduce a suitable global
Cartesian tetrad frame with respect to which the connection vanishes
\begin{equation}\label{gamma0}
  \Gamma^{\a\b}\stareq 0\,.
\end{equation}
Throughout the rest of our paper we will work in a global frame that
obeys Eq.~(\ref{gamma0}). Then all covariant derivatives reduce to
partial derivatives and we can simplify our work considerably. In
particular, the torsion of the Weitzenb\"ock spacetime becomes
\begin{equation}
T^\a:=D\vt^\a=d\vt^\a+\Gamma_\b{}^\a\wedge\vt^\b\stareq d\vt^\a=C^\a\,,
\end{equation}
or, equivalently,
\begin{eqnarray}
  T_{ij}{}^\a&=& \nonumber
  2D_{[i}e_{j]}{}^\a=2\partial_{[i}e_{j]}{}^\a-2
\Gamma_{[ij]}{}^k
  e_k{}^\a \\ &\stareq& 2\partial_{[i}e_{j]}{}^\a =C_{ij}{}^\a\,.
\end{eqnarray}
Accordingly, the {\it gravitational field strength,} in the special
``gauge'' (\ref{gamma0}), is represented by the object of anholonomity
$C^\a$. This concludes the geometrical set-up of a Weitzenb\"ock
spacetime and we will henceforth drop the star over the equal sign.

We now turn to physics in GR$_{||}$, which is determined via the
variation of the action $\d S=0$, where
\begin{equation}\label{action}
S=\int\left({\cal L}_{\text{g}}+{\cal L}_{\text{m}} \right)d^4x\,.
\end{equation}
Here ${\cal L}_{\rm g}$ and ${\cal L}_{\rm m}$ are respectively the
gravitational and matter Lagrangian densities. Out of the field
strength $C^\a$, we can construct a gravitational Lagrangian ${\cal
  L}_{\rm g}$ that is---in analogy with electrodynamics---{\it quadratic}
in the field strength. The explicit form of ${\cal L}_{\rm g}$ can
be left open for the moment. However, we can introduce quite generally
the components of the {\it excitation} ${\cal H}^{ij}{}_\a$ that are
related to those of the field strength by
\begin{equation}{\cal H}^{ij}{}_\a:= -2\frac{\partial {\cal L}_{\rm
    g}}{\partial C_{ij}{}^\a}\,.
\end{equation} 
In terms of differential forms we have $H_\a=\frac 12
H_{ij\a}dx^i\wedge dx^j$, with ${\cal H}^{ij}{}_\a=\frac
12\epsilon^{ijkl}H_{kl\a}$, where $\epsilon^{ijkl}$ is the totally
antisymmetric Levi-Civita symbol with $\epsilon^{0123}=1$. Because of
the presumed quadratic nature of ${\cal L}_{\rm g}$ and the Euler
theorem on homogeneous functions, we find
\begin{equation}\label{Euler}
{\cal L}_{\rm g}=\frac 12 C_{ij}{}^\a\frac{\partial{\cal L}_{\rm g} }
{\partial C_{ij}{}^\a}=-\frac 14{\cal H}^{ij}{}_\a C_{ij}{}^\a\,,
\end{equation}
which expresses the gravitational Lagrangian in terms of the
excitation ${\cal H}^{ij}{}_\a$ and the field strength $C_{ij}{}^\a$
such that ${\cal H}^{ij}{}_\a$ is linear in $C_{ij}{}^\a$.

The field equations of GR$_{||}$ given in Eq.~(\ref{fieldeq}) follow
from the variational principle of stationary action (\ref{action})
with ${\cal T}_\a{}^i:=\d {\cal L}_{\rm m}/\d e_i{}^\a$ as the
source. We note that, in linear approximation, the gravitational
energy-momentum complex (\ref{energy}) will vanish and only the first
term of the left-hand side of Eq.~(\ref{fieldeq})$_1$ will survive.

Having thus established the geometry and the Lagrange-Noether
machinery for the teleparallel theory of gravitation, we turn next to
the explicit choice of the gravitational Lagrangian. General
relativity is recovered via equation (\ref{ten}), which may be written
as
\begin{eqnarray}\label{linconst} {\cal H}^{ij}{}_\a
=
\left\langle\frac{e}{\kappa} g^{k[i}g^{j]l}\eta_{\a\b}
\right\rangle
\mathfrak{C}_{kl}{}^\b\,,
\end{eqnarray}
with the {\it modified} field strength 
\begin{eqnarray}
\mathfrak{C}_{ij}{}^\a :=\frac 12\,
    C_{ij}{}^\a -C^\a{}_{[ij]}+2e_{[i}{}^\a C_{j]\g}{}^\g\,.\label{Cbar}
\end{eqnarray}

In particular, along the same line of thought as in \cite{Muench00},
we wish to consider a nonlocal constitutive law for gravitation. To
prepare the ground, let us introduce the invariant proper
infinitesimal distance $ds$ between two neighboring events in
Weitzenb\"ock spacetime
\begin{equation}\label{lineelement}
ds^2=g_{ij}dx^i\otimes dx^j
\end{equation}
and define a geodesic between two fixed events $P'$ and $P$ to be the
path that is an extremum of the spacetime distance between $P'$ and
$P$,
\begin{equation}\label{vargeo}
\delta\int_{P'}^P ds=0\,.
\end{equation}
This path is given by the geodesic equations
\begin{equation}\label{geodesic}
\frac{d^2x^i}{ds^2}+{\chris{i}{j}{k}}\frac{dx^j}{ds}\frac{dx^k}{ds}=0\,,
\end{equation}
where
\begin{equation}\label{Christoffel}
  {\chris{i}{j}{k}}=\frac 12\,g^{il}\left(g_{lj,k}+g_{lk,j}-g_{jk,l}\right)
\end{equation}
are the Christoffel symbols.

We assume that two causally separated events are connected by a {\it
  unique} timelike or null geodesic; more generally, in the spacetime
region under consideration, there exists a unique geodesic joining
every pair of events. It then proves useful to employ the
{\it world-function} $\Omega$, which denotes half the square of the proper
distance from $P':x'=\xi(\zeta_0)$ to $P:x=\xi(\zeta_1)$ along the
geodesic path $x^i=\xi^i(\zeta)$. That is, we define \cite{Synge}
\begin{equation}\label{world}
  \Omega(x,x')=\frac 12 (\zeta_1-\zeta_0)\int_{\zeta_0}^{\zeta_1}
  g_{ij}\frac{d\xi^i}{d\zeta}\frac{d\xi^j}{d\zeta}d\zeta\,.
\end{equation}
It turns out that $\Omega$ is independent of the affine parameter
$\zeta$; moreover, the integrand in Eq.~(\ref{world}) is constant by
Eq.~(\ref{geodesic}). The main properties of $\Omega(x,x')$ are
summarized in Appendix~\ref{AppB}.

To distinguish coordinate indices that refer to $x$ from those that refer
to $x'$, we henceforth use indices $a, b, c, ...$ at $x$ and $i, j, k,
...$ at $x'$. Thus we define
\begin{equation}\label{world1}
\Omega_a(x,x')=\frac{\partial\Omega}{\partial x^a}\,,\qquad
\Omega_i(x,x')=\frac{\partial\Omega}{\partial x'^i}\,,
\end{equation}
and note that covariant derivatives at $x$ and $x'$ commute for any
bitensor. It follows from the results of Appendix~\ref{AppB} that
\begin{equation}\label{world2}
2\Omega=g^{ab}\Omega_a\Omega_b=g^{ij}\Omega_i\Omega_j\,.
\end{equation}
Differentiating this equation, we find that
$\Omega_{ai}(x,x')=\Omega_{ia}(x,x')$
are smooth dimensionless two-point tensors such that
\begin{equation}\label{2point}
\lim_{x'\rightarrow x}\Omega_{ai}(x,x')=-g_{ai}(x)\,.
\end{equation}
Thus a possible nonlocal generalization of Eq.~(\ref{linconst}) is
given by
\begin{eqnarray}\label{excit2}\hspace{-8pt} {\cal H}^{ab}{}_c(x)&=&
  -\frac{1}{\kappa}\sqrt{-g(x)}\int
  U(x,x')\Omega^{ai}\Omega^{bj}\Omega_{ck}\,{\chi}
  (x,x')\nonumber\\ &&\times\mathfrak{C}_{ij}{}^k
  (x')\sqrt{-g(x')}\,d^4 x'\,,
\end{eqnarray}
where $U(x,x')$ is unity except when $x'$ is in the future of $x$, in
which case $U$ vanishes; in Minkowski spacetime, this means that
events with $x'^0>x^0$ are excluded from the domain of integration in
Eq.~(\ref{excit2}). Moreover, ${\chi}(x,x')$ is a scalar given, for
instance, by
\begin{equation}\label{scalar}
  {\chi}(x,x')=\delta(x-x')+\hat{K}(x,x')\,,
\end{equation}
where the Dirac delta-function $\d(x-x')$ is defined via
\begin{equation}\label{delta}
  \int\delta(x-x')\varphi(x')\sqrt{-g(x')}\,d^4x'=\varphi(x)
\end{equation}
for any smooth function $\varphi(x)$. The scalar kernel
$\hat{K}(x,x')$ in Eq.~(\ref{scalar}) denotes the nonlocal deviation
from a local constitutive law. The main nonlocal constitutive relation
(\ref{excit2}) can therefore be expressed as
\begin{eqnarray}\label{excit3}
  {\cal H}^{ab}{}_c(x)&\!\!=\!\!&\frac{1}{\kappa}\sqrt{-g(x)}\,
  \lbrack\mathfrak{C}^{ab}{}_c(x)
    -\int U(x,x')\Omega^{ai}\Omega^{bj}\Omega_{ck} \,
  \nonumber\\ &&
\times \hat{K}(x,x')
    \mathfrak{C}_{ij}{}^k(x')\sqrt{-g(x')}\,d^4 x'\rbrack\,,
\end{eqnarray}
where the nonlocal deviation from general relativity is made
explicit. More complicated nonlocal constitutive relations are
certainly possible; however, Eq.~(\ref{excit3}) is the simplest one
involving an unknown scalar kernel $\hat{K}$.

We emphasize that the domain of applicability of Eq.~(\ref{excit3}) is
not physically restricted; in particular, a nonlocal gravity theory of
this type is not {\it directly} related to nonlocal special relativity
\cite{BahramAnnalen}. That is, if the whole heuristic machinery of
Einstein's principle of equivalence could be employed in this case,
one would have to conclude---on the basis of nonlocal special
relativity---that gravity should be nonlocal for sufficiently high
gravitational ``accelerations''. However, Einstein's principle of
equivalence cannot be used in this way, as its precise formulation in
GR is the very embodiment of locality. Thus nonlocal gravity is in
this sense {\it decoupled} from nonlocal special relativity.

It is important to note that our ansatz (\ref{excit3}) is {\it
  nonlinear} as well as nonlocal, since, among other things, the
metric tensor (\ref{holmetric}) is quadratic in the gravitational
potentials $e_i{}^\a$. The scalar kernel $\hat{K}(x,x')$ could be
given, for instance, by $\hat{K}^{-1}=\Omega^2+L_0^4$, where $L_0$ is a
constant characteristic length. It could also involve the
Weitzenb\"ock invariants \cite{HehlHeld}
\begin{equation}\label{Winvariants}
C_{ijk}C^{ijk}\,,\quad C_{kji}C^{ijk}\,,\quad C_{ij}{}^j C^{ik}{}_k
\end{equation}
at $x$ and $x'$ as well as scalars formed from the covariant
derivatives of $\Omega(x,x')$. To illustrate the latter, consider, for
example, $\Omega_a$ and $\Omega_i$ in Eq.~(\ref{world1}). The vector
$\Omega^a$ is tangent at $x$ to the geodesic connecting $x'$ to $x$
and the length of $\Omega^a$ is equal to that of this geodesic by
Eq.~(\ref{world2}); in Minkowski spacetime,
$\Omega^a=x^a-x'^{a}$. Similarly, the vector $\Omega^i$ has the same
length, is tangent to the geodesic at $x'$, and is directed away from
$x$; in Minkowski spacetime, $\Omega^i=x'^i-x^{i}$. Hence one can form
coordinate scalars $\Omega^a e_a{}^\a(x)$ and $\Omega^i e_i{}^\a(x')$,
which turn out to be of particular significance in Sec.~V. This brief
discussion indicates that many options indeed exist for generating a
scalar kernel $\hat{K}(x,x')$ in our ansatz (\ref{excit3}). 

In the following section, the nonlocal theory of gravity is treated in
the linear approximation; therefore, Eq.~(\ref{excit3}) will be
employed in linearized form. In particular, $U(x, x')\hat{K}(x, x')$
will be evaluated in Minkowski spacetime, since Weitzenb\"ock
spacetime in the special gauge (\ref{gamma0}) reduces to Minkowski
spacetime when the gravitational potentials $e_i{}^\alpha$ reduce to
$\delta_i^\alpha$. In this {\it limit}, we define 
\begin{equation}\label{KhatK}
K(x, x') = U(x, x')
\hat{K}( x, x')\,.
\end{equation}

\section{Gravitational Field Equations in Linear Approximation}

The gravitational field equations are obtained from the variation of
the action, $\int{\cal L}d^4x$, where ${\cal L}$ is the Lagrangian
density given by the sum of the corresponding quantities for
gravitation ${\cal L}_{\rm g}$ and matter ${\cal L}_{\rm m}$, namely
${\cal L}={\cal L}_{\rm g}+{\cal L}_{\rm m}$, such that
\begin{equation}\label{em}
\frac{\d{\cal L}_{\rm m}}{\d e_i{}^\a}=\sqrt{-g}\,T_\a{}^i\,.
\end{equation}
Here $T_\a{}^i$ is the energy-momentum tensor of matter, so that
$\sqrt{-g}\,T_\a{}^i={\cal T}_\a{}^i$ given in Eq.~(\ref{fieldeq}). The
gravitational part ${\cal L}_{\rm g}$ is given by Eq.~(\ref{Euler}) in
analogy with electrodynamics \cite{Birkbook}. For the nonlocal theory
under consideration, ${\cal H}^{ij}{}_\a$ is now given by
Eq.~(\ref{excit3}); however, to simplify matters, we will work in the
linear approximation. That is, we assume that
\begin{equation}\label{lin1}e^i{}_\a=\d^i_\a-\psi^i{}_\a\,,
\end{equation}
where the nonzero components of $\psi^i{}_\a$ are such that
$|\psi^i{}_\a|\ll 1$. In this approximation, the holonomic coordinate
indices and the anholonomic tetrad indices are indistinguishable. It
then follows from Eq.~(\ref{excit3}) and Appendix~\ref{AppB} that
\begin{eqnarray}\label{Csquare}
{\cal L}_{\rm{g}}&=& \nonumber -\frac{1}{4\kappa} 
\mathfrak{C}^{ij}{}_k(x)C_{ij}{}^k(x)\\ &&-\frac{1}{4\kappa} 
C_{ij}{}^k(x)\int K(x,y)\mathfrak{C}^{ij}{}_k(y)d^4y\,,
\end{eqnarray}
where $K$ is the scalar kernel in the weak-field approximation. A
discussion of action principles in nonlocal field theories is
contained in Edelen \cite{Edelen}. The expressions for $C_{ij}{}^k$
and $\mathfrak{C}^{ij}{}_k$ in terms of $\psi^i{}_j$ can be easily
determined. We find from Eq.~(\ref{lin1}) that
\begin{equation}\label{lin2}
e_i{}^\a=\d_i^\a+\psi^\a{}_i\,,
\end{equation}
so that from (\ref{Omega}),
\begin{equation}
\label{lin3} C_{ij}{}^k=2\psi^k{}_{[j,i]}\,.
\end{equation}
Moreover,
\begin{equation}\label{lin4}
g_{ij}=\eta_{ij}+h_{ij}\,,\qquad h_{ij}=2\psi_{(ij)}\,.
\end{equation}
It follows from (\ref{Cbar}) that
\begin{eqnarray}
\nonumber\mathfrak{C}^{ij}{}_k&=&-\frac
12\left(h_{k\,\;,}^{\,\;i\;\,j}-h_{k\,\;,}^{\,\;j\;\,i}\right)+\psi^{[ij]}{}_{,k}\\
\label{lin5}&&+\, \d^i_k\left(\psi_,{}^j-\psi_{l\;\,,}^{\;\,j\;
    \,l}\right)-\d^j_k\left(\psi_,{}^{\,i}- \psi_{l\;\,,}^{\,\;i\;\,l}\right)\,,
\end{eqnarray}
where $\psi=\eta_{ij}\psi^{ij}$.

Let us first examine the local part of the Lagrangian
(\ref{Csquare}). It follows from
\begin{equation}\label{Llocal}
  \d{\cal L}_{\rm g}(\text{local})=-\frac{1}{4\kappa} \mathfrak{C}^{ij}{}_k
  \left(\d C_{ij}{}^k\right)-\frac{1}{4\kappa}\left(\d\mathfrak{C}^{ij}{}_k
  \right)C_{ij}{}^k
\end{equation}
that the first term in Eq.~(\ref{Llocal}) is in effect given by
\begin{equation}\label{first}-\frac{1}{2\kappa}
  \mathfrak{C}^{ij}{}_{k,j}\left(\d\psi^k{}_i\right)
\end{equation}
via Eq.~(\ref{lin3}) and integration by parts, assuming that
$\d\psi^k{}_i$ vanishes on the boundary, and hence neglecting such
boundary terms in Eq.~(\ref{first}). Applying this same procedure to
the second term in Eq.~(\ref{Llocal}), we find after a detailed
calculation that the result is again given by
\begin{equation}\label{second}
  -\frac{1}{2\kappa}
  \mathfrak{C}^{ij}{}_{k,j}\left(\d\psi^k{}_i\right)\,,
\end{equation}
so that in the {\it absence} of the nonlocal term ($K=0$) the variation of
the action can be expressed in the linear approximation as
\begin{equation}\label{varac}
  \int\left(-\frac{1}{\kappa}\mathfrak{C}^{ij}{}_{k,j}+T_k{}^i\right)
  \left(\d\psi^k{}_i\right)d^4x=0\,.
\end{equation}
Thus for $K=0$, the gravitational field equations are
$\partial_j\mathfrak{C}^{ij}{}_k=\kappa T_k{}^i$; moreover, we prove in
Appendix~\ref{AppC} that these are precisely Einstein's field equations
according to the standard general linear approximation scheme. A
general proof, valid in the non-linear regime, is already contained in
\cite{HehlHeld}. We note that $\partial_iT_k{}^i=0$ is implied by the
field equations.

We now turn to the nonlocal part of the Lagrangian. In this case, the
corresponding two terms in the variation of the gravitational action
are
\begin{eqnarray}\nonumber
&&\hspace{-15pt}\d{\cal L}_{\rm g}(\text{nonlocal})=-\frac{1}{4\kappa}\left[\d
C_{ij}{}^k(x)\right]\int K(x,y)\,\mathfrak{C}^{ij}{}_k(y)d^4y\\
&&\hspace{20pt}-\frac{1}{4\kappa} C_{ij}{}^k(x)\int
K(x,y)\left[\d\mathfrak{C}^{ij}{}_k(y)\right]d^4y\,.\label{both}
\end{eqnarray}
As before, the first term in Eq.~(\ref{both}) is easily shown to lead
to
\begin{equation}\label{firs40}
  -\frac{1}{2\kappa}
  \left[\d\psi^k{}_i(x)\right]\int\frac{\partial K(x,y)}{\partial
    x^j}\mathfrak{C}^{ij}{}_k(y)d^4y\,.
\end{equation}
For the second term in Eq.~(\ref{both}), we can write its contribution
to the variation of the total action as
\begin{equation}\label{totac}
-\frac{1}{4\kappa}\int\int
K(x,y)C_{ij}{}^k(x)\left[\d\mathfrak{C}^{ij}{}_k(y)\right]d^4y\,d^4x\,.
\end{equation}
The domains of integration in the double integral are the same;
therefore, it is possible to switch $x$ and $y$ in Eq.~(\ref{totac}) to
get
\begin{equation}\label{totac1}-\frac{1}{4\kappa}\int\int
K(y,x)C_{ij}{}^k(y)\left[\d\mathfrak{C}^{ij}{}_k(x)\right]d^4y\,d^4x\,.
\end{equation}
Hence the second term in the nonlocal part of the Lagrangian is
equivalent to
\begin{equation}\label{totac2}
-\frac{1}{4\kappa}\left[\d\mathfrak{C}^{ij}{}_k(x)\right]\int
K(y,x)C_{ij}{}^k(y)d^4y\,.
\end{equation}
Now applying to Eq.~(\ref{totac2}) the same procedure we used in the
derivation of Eq.~(\ref{second}), we find after a detailed calculation
that the result is
\begin{equation}\label{second40}
  -\frac{1}{2\kappa}
  \left[\d\psi^k{}_i(x)\right]\int\frac{\partial K(y,x)}{\partial
    x^j}\mathfrak{C}^{ij}{}_k(y)d^4y\,.
\end{equation}

Let us define the symmetric kernel $\overline{K}$ by
\begin{equation}\label{Kover}
\overline{K}(x,y)=\frac 12\left[K(x,y)+K(y,x)\right]\,.
\end{equation}
Then, combining equations (\ref{firs40}) and (\ref{second40}), we
find
\begin{eqnarray}\label{nonloc}
  \d{\cal L}_{\rm
    g}(\text{nonlocal})&=& \nonumber -\frac{1}{\kappa}
\left[\d\psi^k{}_i(x)\right] \\ &&\times\int
  \frac{\partial\overline{K}(x,y)}{\partial
    x^j}\mathfrak{C}^{ij}{}_k(y)d^4y\,.
\end{eqnarray}
This means that the field equations of the nonlocal theory in the
linear approximation are given by
\begin{equation}\label{linapprox}
\partial_j\mathfrak{C}^{ij}{}_k+\int\frac{\partial
\overline{K}(x,y)}{\partial
x^j}\mathfrak{C}^{ij}{}_k(y)d^4y=\kappa T_k{}^i\,.
\end{equation}

We note that $\partial_i T_k{}^i=0$ is still identically satisfied in
the nonlocal theory. For a general symmetric kernel $\overline{K}$,
the energy-momentum tensor $T_{ij}$ is not symmetric in general, since the
nonlocal part in Eq.~(\ref{linapprox}) is not in general
symmetric. This poses no basic difficulty as $T_{ij}$, given in
general by Eq.~(\ref{em}), is not symmetric in general. On the other
hand, the requirement that $T_{ij}$ be symmetric (as in GR, for
instance) would impose restrictions on the kernel $\overline{K}$.

A natural interpretation of the basic nonlocal gravitational field
equations (\ref{linapprox}) can be obtained by considering the
nonlocal term to be an effective source for ``dark matter'' and hence
moving it to the right-hand side of Eq.~(\ref{linapprox}). In this
way, one can recover Einstein's theory but with ``dark matter''. We
explore this possibility in the rest of this paper.

Consider, for instance, the possibility that $\overline{K}$ is an even
function of $x-y$; that is,
\begin{equation}\label{KequalsF}
\overline{K}(x,y)=F(x-y)\,,
\end{equation}
where $F(z)=F(-z)$. Then, ${\partial\overline{K}}/{\partial
  x^j}=-{\partial\overline{K}}/{\partial y^j}$ and Eq.\
(\ref{linapprox}) can be written as
\begin{equation}\label{Tsym}
\partial _j\mathfrak{C}^{ij}{}_k(x)+\int
\overline{K}(x,y)\frac{\partial\mathfrak{C}^{ij}{}_k(y)}{\partial
y^j}d^4y=\kappa T_k{}^i\,,
\end{equation}
which implies, via Appendix~C, that $T_{ki}=T_{ik}$. Here we have
assumed that the derivatives of $\psi^k{}_i$ vanish on the boundary of
the spacetime region under consideration.\bigskip

\section{Nonlocal field equations}

The treatment of the previous section has been based on the assumption
that nonlocal gravitational field equations should be obtained from an
action principle that incorporates the nonlocal constitutive relation
(\ref{excit3}).  Alternatively, this relation could be employed in the
field equations of teleparallelism. The purpose of this section is to
address certain subtleties involved in these possibilities.

\subsection{The ``general'' field equations of teleparallelism}

In electrodynamics the Maxwell equations in (\ref{Max}) can be derived
from electric charge and magnetic flux conservation, see
\cite{Birkbook}. No action principle is necessary. If one considers,
for example, local and linear magnetoelectric matter, the constitutive
law is ${\cal H}^{ij}=\frac 12\chi^{ijkl}F_{kl}$, with a constitutive
tensor density $\chi^{ijkl}=-\chi^{jikl}=-\chi^{ijlk}$ that has 36
independent components. This model can be derived from a Lagrangian,
provided the constitutive tensor obeys additionally the Onsager-type
relation $\chi^{ijkl}=\chi^{klij}$; that is, if 15 of its 36
components vanish. Nevertheless, even if this condition is not
fulfilled, one can still find reasonable applications in physics,
though in this case irreversible processes have to be taken into
account.

The situation is similar for the energy-momentum tensor density of the
electromagnetic field. Starting from the Lorentz force density,
substitution of the inhomogeneous Maxwell equations and partial
integration result in the (canonical) Minkowski energy-momentum
tensor density
\begin{equation}\label{ff}
  {\cal T}_i{}^{j}=-\frac{1}{4}\delta_i^j F_{kl}{{\cal
    H}}^{kl}+F_{ik}{{\cal H}}^{jk}\,.
\end{equation}
Apparently no Lagrangian is necessary for the derivation of this
expression \cite{Minkowski,Birkbook}.

The gravitational field equations of the translational gauge theory,
as formulated in Eq.~(\ref{fieldeq}), should also have general
validity within this framework, independently of the existence of the
Lagrangian. For the homogeneous equations (\ref{fieldeq})$_2$ this is
apparent since they represent just the first Bianchi identities in a
Weitzenb\"ock spacetime. For the inhomogeneous equations
(\ref{fieldeq})$_1$ this can be seen as follows: We start with the
{\it energy-momentum conservation law for matter} in the linear
approximation $\partial_i {\cal T}_\a{}^i\sim 0$. We can ``solve'' it
by $ {\cal T}_\a{}^i=\partial_j\widetilde{\cal
  H}^{ij}{}_\a-\widetilde{\cal E}_\a{}^i$ and $\widetilde{\cal
  H}^{ij}{}_\a=-\widetilde{\cal H}^{ji}{}_\a$, with some unknown
correction term $\widetilde{\cal E}_\a{}^i$. These are already the
inhomogeneous field equations. We just have to determine
$\widetilde{\cal E}_\a{}^i$ explicitly. Clearly it adds to $ {\cal
  T}_\a{}^i$, namely $\partial_j\widetilde{\cal H}^{ij}{}_\a= {\cal
  T}_\a{}^i+\widetilde{\cal E}_\a{}^i$; then, it is natural to
interpret $\widetilde{\cal E}_\a{}^i$ as the energy-momentum tensor
density of the gravitational field. In analogy with Eq.~(\ref{ff}) we
expect it to have the form of Eq.~(\ref{energy}); we now drop the
tildes from $\widetilde{\cal E}_\a{}^i$ and $\widetilde{\cal
  H}^{ij}{}_\a$. Therefore the inhomogeneous field equations extracted
from the energy-momentum conservation law of matter in this heuristic
manner have exactly the same form as (\ref{fieldeq})$_1$ with
(\ref{energy}). Thus we may postulate (\ref{fieldeq}) as the general
field equations valid independently of the existence of a Lagrangian.

By contrast, if we have a Lagrangian, the field equations turn out to
be \cite{HehlHeld}
\begin{equation}\label{energy1}
  \partial_j{\cal H}^{ij}{}_\a- \left(e^i{}_\a{\cal L}_{\rm g} 
    + C_{\a k}{}^\b {\cal H}^{ik}{}_\b\right)={\cal T}_\a{}^i\,.
\end{equation}
If ${\cal L}_{\rm g}=-\frac 14 C_{ ij}{}^\a {\cal H}^{ij}{}_\a$ is
substituted into (\ref{energy1}), we recover (\ref{fieldeq})$_1$.
Consequently, we have shown that the ``general'' field equations
(\ref{fieldeq})$_1$ with (\ref{energy}) are correct if a Lagrangian
exists. Accordingly, they are certainly one consistent and reasonable
generalization of the Lagrangian-based equations (\ref{energy1}).

\subsection{Ambiguity in the field equations}

We now wish to address a certain ambiguity that is encountered in
implementing a nonlocal constitutive law as in our present work. A
{\it local} constitutive law relating the excitation ${\cal
  H}^{ij}{}_k$ to the gravitational field stregth $C^{ij}{}_k$ can be
used in the derivation of the field equations of the theory in either
of two {\it equivalent} ways: it could be employed in the Lagrangian
(\ref{Euler}) that then results, via the variational principle of
stationary action, in the desired field equations or, alternatively, it
could be directly substituted in the field equations
(\ref{fieldeq})$_1$. These two methods in general produce different
results for a {\it nonlocal} constitutive law, however.  It is
interesting to illustrate this point for the case at hand in the
linear approximation. In this regime, we have the linear constitutive
law
\begin{eqnarray}\label{option1}
\kappa {\cal H}^{ij}{}_\alpha (x)= \mathfrak{C}^{ij}{}_\alpha (x)
                + \int K(x, y) \mathfrak{C}^{ij}{}_\alpha (y) d^4y\,,
\end{eqnarray}
which, when inserted in Eq.~(\ref{Euler}) results, as described in
detail in Sec.\ III, in the nonlocal field equations
(\ref{linapprox}). Alternatively, however, we can equally well
substitute the constitutive relation (\ref{option1}) in equation
(\ref{fieldeq})$_1$, which reduces in the linear regime to the new
nonlocal field equations
\begin{equation}\label{option2}
\partial_j\mathfrak{C}^{ij}{}_k+\int\frac{\partial
{K}(x,y)}{\partial
x^j}\mathfrak{C}^{ij}{}_k(y)d^4y=\kappa T_k{}^i\,.
\end{equation}
The two nonlocal field equations only differ in their kernels:
Eq.~(\ref{linapprox}) involves $\overline{K}$, while Eq.~(\ref{option2})
involves $K$. The question of whether such nonlocal equations as
(\ref{option2}) can be derived from a variational principle is beyond
the scope of our investigation; for a related study in connection with
acceleration-induced nonlocality see \cite{Chicone07}.

In a similar way as in Sec.\ III, one can recover a direct nonlocal
generalization of Einstein's theory with a symmetric energy-momentum
tensor of matter by assuming that $K(x, y)$ is a function of $x-y$. In
this case, Eq.~(\ref{option2}) can be written as
\begin{equation}\label{option3}
\partial _j\mathfrak{C}^{ij}{}_k(x)+\int
{K}(x,y)\frac{\partial\mathfrak{C}^{ij}{}_k(y)}{\partial
y^j}d^4y=\kappa T_k{}^i\,,
\end{equation}
which should be compared and contrasted with Eq.~(\ref{Tsym}). These
inequivalent nonlocal field equations both admit a natural
interpretation in terms of dark matter as described in detail in the
following section; however, Eq.~(\ref{option3}) has an advantage over
Eq.~(\ref{Tsym}) in that it involves a causal kernel $K$, while in the
symmetric kernel $\overline{K}$ of Eq.~(\ref{Tsym}) past and future are
treated in the same manner.

\section{``Dark Matter''}

Let us now consider an important consequence of our nonlocal equations
for the gravitational field. To keep our discussion general, we start
with an equation that has the same form as Eqs.~(\ref{linapprox}) and
(\ref{option2}), except for a kernel ${\cal K}$, which we take to be
equal to $\overline{K}$ or $K$, respectively, depending on whether one
adopts the Lagrangian-based approach of Sec.\ III or the more direct
approach of Sec.\ IV. We henceforth assume, for simplicity, that in
the former case $\overline{K}(x, y)$ is an even function of $x-y$ and
in the latter case $K(x, y)$ is simply a function of $x-y$.  Then, the
arguments of the previous sections imply that the nonlocal modification
of Einstein's gravitational field equations for a symmetric
energy-momentum tensor $T_{ij}$ may be expressed as
\begin{equation}\label{integraleq}
  G_{ij}(x)+\int{\cal K}(x,y)G_{ij}(y)d^4y=\kappa T_{ij}(x)\,,
\end{equation}
where $G_{ij}$ is given by
\begin{equation}\label{feq1}
  G^i{}_j=\partial_k\mathfrak{C}^{ik}{}_j
\end{equation}
and represents Einstein's tensor in the linear approximation.

The nonlocal term in equation (\ref{integraleq}) can be interpreted in
terms of an {\it effective} energy-momentum tensor for ``dark matter''
by writing Eq.~(\ref{integraleq}) as 
\begin{equation}\label{feqdark}
G_{ij}(x)=\kappa\left[T_{ij}(x)+\mathfrak{T}_{ij}(x) \right]\,,
\end{equation}
where the ``dark matter'' component is given by the symmetric
energy-momentum tensor
\begin{equation}\label{dark1}
\mathfrak{T}_{ij}(x)=-\frac 1\kappa\int{\cal K}(x,y)G_{ij}(y)d^4y\,.
\end{equation}

Equation (\ref{dark1}) may be written in a more transparent form by
making use of the methods of Appendix~\ref{AppD}. That is, suppose
that the integral equation (\ref{integraleq}) can be solved by the
method of successive substitutions (cf.\ Appendix~\ref{AppD}) and that
the resulting infinite series is uniformly convergent. Then, as shown
in Appendix~D, it is possible to introduce a reciprocal scalar kernel
${\cal R}$ such that
\begin{equation}\label{feq2}
  G_{ij}(x)=\kappa T_{ij}(x)+\kappa\int{\cal R}(x,y)T_{ij}(y)d^4y\,.
\end{equation}
Thus the ``dark matter'' component in equation (\ref{feqdark}) is
given by
\begin{equation}\label{dark2}
\mathfrak{T}_{ij}(x)=\int{\cal R}(x,y)T_{ij}(y) d^4y\,,
\end{equation}
which is the integral transform of matter $T_{ij}$ by the kernel
${\cal R}(x,y)$. It is clear from Eq.~(\ref{dark2}) that in our
model, ``dark matter'' should be quite similar in its characteristics
to actual matter; for instance, for dust, the corresponding ``dark
matter'' is pressure-free, while for radiation with $T_k{}^k = 0$, the
``dark'' energy-momentum tensor $\mathfrak{T}_{ij}$ is traceless as well.

The reciprocal kernel ${\cal R}$ is given as an infinite series in
terms of iterated kernels constructed from ${\cal K}(x,y)$; that is,
\begin{equation}\label{kernel1}
-{\cal R}(x,y)=\sum_{n=1}^\infty{\cal K}_n(x,y)\,.
\end{equation}
To ensure causality, it is useful to assume further that
\begin{equation}\label{product}
{\cal K}(x,y)=\d (x^0-y^0)P(\mathbf{x},\mathbf{y})\,,
\end{equation}
where $P$ is a function of $\mathbf{x}-\mathbf{y}$. Then, all
iterated kernels as well as the reciprocal kernel are also of this
general form; indeed, 
\begin{equation}\label{kernel2}
{\cal K}_n(x,y)=\d (x^0-y^0)P_n(\mathbf{x},\mathbf{y})\,,
\end{equation}
where $P_1(\mathbf{x},\mathbf{y})=P(\mathbf{x},\mathbf{y})$ and
\begin{equation}\label{kernel3}
  P_{n+1}(\mathbf{x},\mathbf{y})=-\int
  P(\mathbf{x},\mathbf{z}) P_n(\mathbf{z},\mathbf{y}) d^3z\,.
\end{equation}
Therefore,
\begin{equation}\label{overline}
{\cal R}(x,y)=\d (x^0-y^0)Q(\mathbf{x},\mathbf{y})\,,
\end{equation}
where $Q$ is reciprocal to $P$,
\begin{equation}\label{quu}
-Q(\bfx,\bfy)=\sum\limits_{n=1}^\infty P_n(\bfx,\bfy)\,.
\end{equation}
Moreover, $P_n$ and $Q$ are functions of $\mathbf{x}-\mathbf{y}$,
since the integration in Eq.~(\ref{kernel3}) extends over the entire
three-dimensional Euclidean space. Thus ``dark matter'' is the
convolution of matter and the reciprocal kernel in this case. The
mathematical implications of this fact are treated in Appendix E.

It is interesting to consider the confrontation of our nonlocal theory
in the linear approximation, represented by equations
(\ref{integraleq})-(\ref{dark2}), with experimental data. Let us first
determine the Newtonian limit of our nonlocal gravity. This follows
directly from Eq.~(\ref{feqdark}) in the standard manner and we find
the Poisson equation
\begin{equation}\label{firstorder}
\nabla^2  \Phi=4\pi G(\rho+\rhod)\,,
\end{equation}
where $\Phi$ is the Newtonian potential. Here $\rho$ is the density of
matter and $\rhod$,
\begin{equation}\label{darkdens}
  \rhod(t,\bfx)=\int Q(\bfx,
  \bfy)\rho(t,\bfy) d^3y\,,
\end{equation}
is the corresponding density of ``dark matter''. We note that in the
general linear approximation of our theory the kernel ${\cal
  K}(x,y)$ and its reciprocal ${\cal R}(x,y)$ are given
functions in Minkowski spacetime and are thus {\it independent} of any
particular physical system. The solar-system tests of
general relativity imply that our ``dark matter'' must be a rather
small fraction of the actual matter in the solar system. This can be
simply arranged with a suitable choice of the universal reciprocal
kernel $Q$.

It is interesting to illustrate these considerations in the case of
the problem of dark matter in spiral galaxies
\cite{Rubin,Roberts,Sofue,Carignan,Jacob2004,Jacob2006,Bruneton2008}. Imagine,
for instance, the circular motion of stars in the disk of a spiral
galaxy. Beyond the galactic bulge, the Newtonian acceleration of
gravity for each star at radius $|\mathbf{x}|$ is given by
$v_0^{\,2}/|\mathbf{x}|$ toward the center of the galaxy. Here
$v_0$ is the constant ``asymptotic'' speed of stars in
accordance with the observed rotation curves of spiral galaxies. It
follows from Poisson's equation that the effective density of dark
matter is essentially given by $v_0^{\,2}/(4\pi
G|\mathbf{x}|^2)$. Comparing this result with equation
(\ref{darkdens}) and setting
\begin{equation}\label{em1}
\rho(t,\mathbf{y})=M\d (\mathbf{y})\,,\qquad
\rhod(t,\mathbf{x})=\frac{v_0^{\,2}}{4\pi G|\mathbf{x}|^2}\,,
\end{equation}
where $M$ is the effective mass of the galaxy and the dimensions of
the galactic bulge have been ignored, we find
\begin{equation}\label{Qx}
Q(\bfx,\mathbf{0})=\frac{v_0^{\,2}}{4\pi GM|\mathbf{x}|^2}\,.
\end{equation}
The reciprocal kernel $Q$ is a function of $\bfx-\bfy$; therefore, it
follows from Eq.~(\ref{Qx}) that 
\begin{equation}\label{Qxy}
Q(\bfx,\bfy)=\frac{1}{4\pi\lambda}\, \frac{1}{|\bfx-\bfy|^2}
\end{equation}
with a universal length parameter $\lambda=GM/v_0^{\,2}$. Thus taking due
account of the observed rotation curves of spiral galaxies,
Eqs.~(\ref{firstorder}) and (\ref{darkdens}) with kernel (\ref{Qxy})
imply
\begin{equation}\label{79}
  \nabla^2  \Phi=4\pi G\left[\rho(t,\bfx)+\frac{1}{4\pi\lambda}
    \int\frac{\rho(t,\bfy)d^3y}{|\bfx-\bfy|^2}\right]\,.
\end{equation}
It is important to point out in passing a defect of the specific form
of Eq.~(\ref{Qxy}): the total amount of ``dark matter'' associated
with a nonzero matter density $\rho$ is infinite (see Appendix E);
therefore, kernel (\ref{Qxy}) is too simple to be quite adequate for
the task at hand.

For a point mass $m$ with $\rho(t,\bfx)=m\d(\bfx)$, the Newtonian
potential given by Eq.~(\ref{79}) can be expressed as
\begin{equation}\label{80}
  \Phi(t,\bfx)=-\frac{Gm}{|\bfx|}+\frac{Gm}{\lambda}
  \ln\left(\frac{|\bfx|}{\lambda}\right)\,.
\end{equation}
The observational data for spiral galaxies indicate that $\lambda$ is
of the order of a kpc, so that the logarithmic term in
Eq.~(\ref{80}) is essentially negligible in the solar
system. Moreover, the universality of the kernel in the linear
approximation implies, via $\lambda=GM/v_0^{\,2}$, that for spiral
galaxies $M\propto v_0^{\,2}$, since $\lambda$ should be independent of
any particular physical system.

It is remarkable that in the simple considerations regarding
Eqs.~(\ref{79}) and (\ref{80})---based on our linear approximation
scheme---we have recovered the significant proposal put forward by
Tohline and further developed by Kuhn et al.\
\cite{TohlineKuhn,Jacob1988} to solve the dark matter problem by a
natural modification of the Newtonian law of gravitation. An
interesting discussion of the Tohline-Kuhn scheme is contained in the
review paper of Bekenstein \cite{Jacob1988}. Despite various
successes, the main drawback of this approach appears to be the
violation of the Tully-Fisher relation, which implies that $M\propto
v_0^{\,4}$ \cite{TohlineKuhn,Jacob1988}. To agree with the empirical
Tully-Fisher law, it seems necessary to go beyond the linear
approximation scheme.

Within the general framework of this work, but going beyond the linear
approximation as well as the specific constitutive model employed thus
far, it may be possible in principle to have a relation of the general
form of Eq.~(\ref{dark2}) with a kernel that is highly dependent upon
the particular system under consideration.  This could provide a
natural way to interpret the observational evidence for dark matter as
the nonlocal manifestation of the gravitational interaction. It
remains to elucidate the physical origin of the constitutive kernel
that has been the starting point of our investigation.\bigskip

\section{Conclusion}

To develop a nonlocal generalization of GR, it proves useful to
approach Einstein's theory via its equivalent within teleparallelism
gravity, namely, GR$_{||}$. Therefore, we work in Weitzenb\"ock
spacetime with a {\it tetrad} field $e_i{}^\alpha$ and its dual
$e^j{}_\beta$, a {\it metric} $g_{ij}$, and a flat {\it connection}
$\Gamma_i{}^{\alpha \beta} = - \Gamma_i{}^{\beta \alpha}$ that is
chosen to vanish globally ($\Gamma_i{}^{\alpha \beta} = 0$). The
gravitational field strength is given by $C_{ij}{}^\alpha$ and the
modified field strength by $\mathfrak{C}_{ij}{}^\alpha$. In this
framework, which is capable of nonlocal generalization, GR$_{||}$
corresponds to a specific gravitational Lagrangian. Working with a
nonlocal ``constitutive'' kernel $K(x, x')$ in the linear
approximation, we construct an explicit nonlocal generalization of
Einstein's theory of gravitation that is consistent with
causality. This theory can be reformulated as linearized general
relativity but with ``dark matter'', which mimics the contribution of
nonlocal gravity. We find that the effective energy-momentum tensor of
``dark matter'' is simply the integral transform of the
energy-momentum tensor of matter.

The application of our nonlocal model in the linear approximation to
the dark-matter problem in spiral galaxies is in conflict with the
empirical Tully-Fisher relation. It is possible that the situation can
be significantly improved with a suitable choice for the kernel in the
nonlinear regime. However, a more basic theory is needed to determine
the nonlocal constitutive kernel from first principles. This is a task
for the future. \bigskip

{\it Acknowledgments.}  We are grateful to Jacob Bekenstein
(Jerusalem), Yakov Itin (Jerusalem), Jos\'e Maluf (Brasilia), Eckehard
Mielke (Mexico City), and Yuri Obukhov (Moscow) for helpful comments.

\appendix

\section{Translational gauge theory in exterior calculus}\label{AppA}

The coframe 1-form $\vt^\a= e_i{}^\a dx^i$ represents the
gravitational potential; here $i,j,...=0,1,2,3$ are (holonomic)
coordinate indices and $\a,\b,...=\hat{0},\hat{1},\hat{2},\hat{3}$
are (anholonomic) frame indices; the frame $e_\b=e^j{}_\b\partial_j$ is
dual to the coframe, that is, $e_i{}^\a e^i{}_\b=\d^\a_\b$ and
$e_i{}^\a e^j{}_\a=\d_i^j$. The field
strength of gravitation is the object of anholonomity 2-form
$ C^\a:=d\vt^\a$.

Spacetime is described by a Weitzenb\"ock geometry with a teleparallel
connection 1-form $\Gamma_\a{}^\b=\Gamma_{i\a}{}^\b dx^i$ and with the
curvature 2-form
\begin{equation}\label{tele}
  R_\a{}^\b
  :=d\Gamma_\a{}^\b-\Gamma_\a{}^\g\wedge \Gamma_\g{}^\b 
=\frac 12 R_{ij\a}{}^\b dx^i\wedge dx^j
\end{equation}
that vanishes
\begin{equation}\label{tele1}
 R_\a{}^\b=0\,.
\end{equation}
 We decompose the connection into a Riemannian (Levi-Civita) part and
 the contortion,
\begin{equation}\label{decomp}
  \Gamma_\a{}^\b=\widetilde{\Gamma}_\a{}^\b-K_\a{}^\b\,,
\end{equation}
with the following definitions of the torsion 2-form and the
contortion 1-form, respectively,
\begin{equation}\label{tortion}
  T^\a:=D\vt^\a=d\vt^\a+\Gamma_\b{}^\a\wedge\vt^\b\,,\;
  T^\a=:K^\a{}_\b\wedge\vt^\b\,.
\end{equation}

As in a Euclidean space, one can likewise choose in a Weitzenb\"ock
space suitable frames such that the connection vanishes everywhere:
\begin{equation}\label{gauge}
\Gamma_\a{}^\b=\Gamma_{i\a}{}^\b dx^i\stackrel{*}{=}0\,.
\end{equation}
Eq.~(\ref{gauge}) should be understood as a choice of a certain gauge,
which is possible since the curvature $R_\a{}^\b$ vanishes
everywhere. Many calculations are simplified in this gauge.

The Lagrangian 4-form is the sum of a gravitational part and a matter part
\begin{equation}\label{Lagrangian}
  L=L_{\text{g}}(\vt^\a,C^a)+ L_{\text{m}}(\psi, d\psi,\vt^a)\,.
\end{equation}
The field equation reads
\begin{equation}\label{fieldeq1}
  dH_\a-E_\a=\frac{\d L_{\text{m}} }{\d\vt^\a}=\Sigma_\a\,.
\end{equation}
Here we have the 2-form $H_\a:=-\partial L_{\text{g}}/\partial C^\a
=\frac 12 H_{ij\a} dx^i\wedge dx^j$ as the excitation and the 3-form
$\Sigma_\a$ as the energy-momentum density of matter, whereas the
3-form $E_\a$ represents the energy-momentum density of the
gravitational field \cite{PRs}
\begin{equation}\label{gaugeenergy}
  E_\a:=e_\a\rfloor L_{\text{g}}+ \left( e_\a\rfloor
    C^\b\right)\wedge H_\b\,.
\end{equation}

Equations (\ref{fieldeq1}) with (\ref{gaugeenergy}) represent the
Lagrangian-based form of the field equations. If we express the
gravitational energy-momentum 3-form $E_\a$ in terms of a Minkowski
type current [or, equivalently, if we substitute $L_{\rm g}=-\frac 12
C^\a\wedge{\cal H}_\a$ into Eq.~(\ref{gaugeenergy})], we find the
general field equations
\begin{equation}
 dH_\a-{\frac 1 2}\left[H_\b\wedge (e_\alpha\rfloor C^\b)
-C^\b\wedge(e_\alpha\rfloor
    H_\b) \right]\stareq\Sigma_\a\,.
\end{equation} 
This is the exterior-form version of the field equations
(\ref{fieldeq})$_1$ with Eq.~(\ref{energy}).

Usually $H_\a$ is, like in electrodynamics, linear in the field
strength
\begin{equation}\label{lin}
  H_\a=\frac{g_{\a\b}}{\kappa}\,^\star(a_1\,^{(1)}C^\b+a_2
  \,^{(2)}C^\b+
  a_3\,^{(3)}C^\b)\,.
\end{equation}
Here $^{(I)}C^\b$ are the different irreducible pieces of $C^\b$ and
$^\star$ represents the Hodge star.  The ``Einsteinian'' choice for
$^{(I)}C^\b$,  $I$ = 1, 2, 3, turns out to be
\begin{equation}\label{Echoice}
a_1=-1\,,\quad a_2=2\,,\quad a_3=\frac 12\,.
\end{equation}
In this case, Eq.~(\ref{lin}) is distinguished from all
linear Lagrangians in that it becomes {\it locally} Lorentz covariant.

\section{Properties of the world-function $\Omega$}\label{AppB}

Consider a variation of Eq.~(\ref{world}) that changes the endpoints, then
\begin{equation}\label{deltaOmega}
\delta\Omega(x,x')=(\zeta_1-\zeta_0)\left[g_{ij}\frac{d\xi^j}{d\zeta}
\delta\xi^i\right]_{\zeta_0}^{\zeta_1}\,.
\end{equation}
On the other hand,
\begin{equation}\label{d01}
\d \Omega=\frac{\partial\Omega}{\partial x^a}\d x^a+
\frac{\partial\Omega}{\partial x'^{i}}\d x'^{i}\,,
\end{equation}
so that
\begin{eqnarray}\label{d02}
  \frac{\partial\Omega}{\partial
    x^a}&=& \nonumber (\zeta_1-\zeta_0)g_{ab}(x)\frac{dx^b}
{d\zeta}\,,\\\frac{\partial\Omega}{\partial x'^{i}}&=&-
(\zeta_1-\zeta_0)g_{ij}(x')\frac{dx'^{j}}{d\zeta}\,.
\end{eqnarray}
It is possible to see from Eq.~(\ref{geodesic}) that the integrand in
Eq.~(\ref{world}) is indeed constant; therefore,
\begin{eqnarray}\label{Omega1}
  \Omega(x,x')&=& \nonumber \frac 12(\zeta_1-\zeta_0)^2 
g_{ab}(x)\frac{dx^a}{d\zeta}
  \frac{dx^b}{d\zeta}\\ &=&\frac 12(\zeta_1-\zeta_0)^2
  g_{ij}(x')\frac{dx'^i}
  {d\zeta}  \frac{dx'^j}{d\zeta}\,.
\end{eqnarray}
It follows from Eqs.~(\ref{d02})-(\ref{Omega1}) that Eq.~(\ref{world2})
is satisfied; moreover, $\Omega=0$ for a null geodesic, $\Omega=\frac
12\,\tau^2$ for a timelike geodesic of length $\tau$ and
$\Omega=-\frac 12\,\sigma^2$ for a spacelike geodesic of length $\sigma$.

In Minkowski spacetime $\Omega$ is given by
\begin{equation}\label{OmegaMink1}
\Omega(x,x')=\frac 12\eta_{ij}(x'^i-x^i)(x'^j-x^j)\,.
\end{equation}
According to our convention, $\eta_{ij}=\text{diag}(1,-1,-1,-1)$;
hence,
\begin{equation}\label{OmegaMink2}
  \Omega(x,x')=\frac 12\left[(t'-t)^2-(\mathbf{x'}-\mathbf{x})^2 \right]\,.
\end{equation}
In this case, we find that
\begin{equation}\label{OmegaMink3}
\Omega_{ai}=\frac{\partial^2\Omega(x,x')}{\partial x^a\partial x'^i}=
-\eta_{ai}\,,
\end{equation}
while
\begin{equation}\label{OmegaMink4}
  \Omega_{ab}=\frac{\partial^2\Omega}{\partial x^a\partial x^b}=
  \eta_{ab}\,,\quad\Omega_{ij}=\frac{\partial^2\Omega}{\partial x'^i
\partial x'^j}=\eta_{ij}\,.
\end{equation}

\section{Symmetry of the tensor
  $\partial_j\mathfrak{C}_{i\;k}^{\;j}$} \label{AppC}

The purpose of this appendix is to show that
$\partial_j\mathfrak{C}_{i\;k}^{\;j}$ is a symmetric tensor, so that
in the field equations
\begin{equation}
\label{fieldeqsym}
\partial_j\mathfrak{C}^{ij}{}_k=\kappa T_k{}^i
\end{equation}
the energy-momentum tensor of the source is symmetric
($T_{ik}=T_{ki}$). In fact Eq.~(\ref{fieldeqsym}) is identical to
Einstein's field equations in the linear approximation.

It follows from Eq.~(\ref{lin5}) that
\begin{eqnarray}
\partial_j\mathfrak{C}^{ij}{}_k&=&\nonumber-\frac
12\left(\square\, 
 h_k{}^i-h_{kj,}{}^{ij}\right)+\frac 12
h^{ij}{}_{,kj}\\
\label{feqlin}&&-\frac 12\d^i_kh^{jl}{}_{,jl}
+\d^i_k\,\square\,\psi-\psi_{,\,\;k}^{\,\;i}\,,
\end{eqnarray}
where we have used the relation $2\psi^{jl}{}_{,\,jl}=h^{jl}{}_{,\,jl}$.
Define the trace-reversed gravitational potentials
$\overline{h}_{ik}$ as
\begin{equation}\label{hbar}
\overline{h}_{ik}=h_{ik}-\frac 12\eta_{ik}h\,,
\end{equation} 
where $h=\eta_{ij}h^{ij}=2\psi$. Then, replacing $h_{ij}$ by
$\overline{h}_{ij}+\eta_{ij}\psi$ everywhere in Eq.~(\ref{feqlin})
results in
\begin{equation}\label{feqlin1}
\partial_j\mathfrak{C}_{i\;k}^{\;j}=-\frac
12\square\, 
\overline{h}_{ki}+\frac 12\overline{h}_{k\,\;,ij}^{\,\;j}+\frac
12\overline{h}_{i\;,kj}^{\,\;j}-\frac
12\eta_{ki}\overline{h}^{jl}{}_{\,\,,jl}\,,
\end{equation}
which is manifestly symmetric in $i$ and $k$. From
Eqs.~(\ref{fieldeqsym}) and (\ref{feqlin1}), we get
\begin{equation}\label{feqlin2}-\square\,\overline{h}_{ik}
  +\overline{h}_{i\,\;,jk}^{\,\;j}+\overline{h}_{k\,\;,ji}^{\,\;j}
-\eta_{ik}\overline{h}^{jl}{}_{,jl}=2\kappa T_{ik}\,.
\end{equation} These are exactly the same as Einstein's field
equations in the linear approximation.

\section{Liouville-Neumann method of successive
  substitutions}\label{AppD}

Consider a linear integral equation of the second kind given by
\begin{equation}\label{phi1}
\phi(x)=f(x)+\int_a^b k(x,y)\phi(y)dy\,,
\end{equation}
where $a$ and $b$ are constants. We seek a solution of this Fredholm
equation by the method of successive substitutions. That is, we
replace $\phi$ in the integrand by its value given by equation
(\ref{phi1}). Repeating this process eventually leads to an infinite
series of the type
\begin{eqnarray}\nonumber
\phi(x)&=& f(x)+\int_a^b k(x,y)f(y)dy\\
&&\hspace{-8pt}+\int_a^b\int_a^bk(x,z)k(z,y)f(y)dy\,dz+\cdots\,.\label{phi2}
\end{eqnarray}
If this series is uniformly convergent, we have a solution of the
integral equation (\ref{phi1}). This solution is unique in the space
of real continuous functions on the interval $[a,b]$; a generalization
of this result to the space of square-integrable functions is
contained in Tricomi \cite{Tricomi}.

Let us define the iterated kernels $k_n,\,n=1,2,\dots,$ by
\begin{eqnarray}\label{ker1}
  k_1(x,y)&=& \nonumber k(x,y)\,,\\ k_{n+1}(x,y)&=&\int_a^b k(x,z)k_n(z,y)dz\,.
\end{eqnarray}
These functions occur in the infinite series of Eq.~(\ref{phi2}). We
define the {\it reciprocal} kernel $r(x,y)$ such that
\begin{equation}\label{ker2}
-r(x,y)=\sum_{n=1}^\infty k_n(x,y)\,.
\end{equation}
Then, Eq.~(\ref{phi2}) can be written as
\begin{equation}\label{phi3}
f(x)=\phi(x)+\int_a^b r(x,y)f(y)dy\,.
\end{equation}
It is clear from equations (\ref{phi1}) and (\ref{phi3}) that the
kernels $k$ and $r$ are reciprocal of each other.

Let us note here some properties of these kernels. It follows from
Eq.~(\ref{ker1}) that
\begin{equation}\label{ker2a}
k_{n+p}(x,y)=\int_a^b k_n(x,z)k_p(z,y) dz\,,
\end{equation}
where $p=1,2,\dots\;$. Using Eqs.~(\ref{ker1}) and (\ref{ker2}) we find that
\begin{eqnarray}\nonumber
k(x,y)+r(x,y)&=& \int_a^b k(x,z) r(z,y) dz\\
\label{ker3}&=& \int_a^b r(x,z) k(z,y) dz\,.
\end{eqnarray}
Moreover, if $k$ is symmetric, $k(x,y)=k(y,x)$, then all iterated
kernels as well as $r(x,y)$ are likewise symmetric.

\section{Convolution kernels}\label{AppE}

Suppose that the spatial kernels in Sec.~V are such that
\begin{equation}\label{peequu}
P(\bfx,\bfy)=p(\bfx-\bfy)\,,\quad Q(\bfx,\bfy)=q(\bfx-\bfy)\,.
\end{equation}
The purpose of this appendix is to point out the consequences of the
convolution theorem for a system of density $\rho$ such that the
density of ``dark matter'' $\rhod$ is a convolution of $\rho$ and
$q$. In general, the spatial kernels $p$ and $q$ could depend on the
characteristics of the particular system under consideration.  The
main results of interest here are Eqs.~(\ref{product}) to
(\ref{darkdens}). Let us note that the particular solution of
Poisson's equation (\ref{firstorder}) is given by
\begin{equation}\label{PoissonSol}
  \Phi(t,\bfx)=-G\int\frac{\rho(t,\bfy)+\rhod(t,\bfy)}
  {|\bfx-\bfy|}d^3y\,,
\end{equation}
provided $\Phi$ vanishes sufficiently fast at spatial infinity. It
follows from Eq.~(\ref{darkdens}) that this
Newtonian potential can also be expressed as
\begin{equation}\label{res}
  \Phi(t,\bfx)=-G\int S(\bfx-\bfy)\rho(t,\bfy)d^3y\,,
\end{equation}
where
\begin{equation}\label{ess}
S(\bfr)=\frac{1}{|\bfr|}+\int\frac{q(\bfz)d^3z}{|\bfr-\bfz|}\,.
\end{equation}

Let us assume that the functions of interest here can be expressed as
Fourier integrals; that is, for a function $f$,
\begin{equation}\label{Fourier}
 \nu f(\mathbf{r})=\int\hat{f}(\mathbf{k}) 
e^{i\mathbf{k}\cdot\mathbf{r}}d^3k\,,
\end{equation}
where $\hat{f}(\bfk)$ is given by
\begin{equation}\label{Fourier1}
\nu \hat{f}(\mathbf{k})=\int{f}(\mathbf{r}) 
e^{-i\mathbf{k}\cdot\mathbf{r}}d^3r\,.
\end{equation}
For the sake of simplicity, we have introduced here a constant
parameter $\nu$,
\begin{equation}\label{lambda}
\nu=(2\pi)^{3/2}\,.
\end{equation}
We are interested in Fourier integral transforms of {\it
  real} functions; therefore,
$\hat{f}^*(\mathbf{k})=\hat{f}(-\mathbf{k})$.

It follows from the convolution theorem for Fourier integrals that
Eq.~(\ref{darkdens}) implies
\begin{equation}\label{convolution}
  \hat{\rhod}(\bfk)=
  \nu\hat{\rho}(\bfk)\hat{q}(\bfk)\,.   
\end{equation}
Here $\hat{q}(\bfk)$ is a dimensionless function such that
\begin{equation}\label{mass}
\nu\hat{q}(\mathbf{0})=\frac{M_{\rm D}}{M}\,,
\end{equation}
where $M=\nu\hat{\rho}(\mathbf{0})$ is the mass of the system
under consideration and $M_{\rm D}=\nu \hat{\rhod}(\mathbf{0})$ is
the corresponding ``dark'' mass. Consider, for instance, kernel
(\ref{Qxy}) that is associated with the discussion of the rotation
curves of spiral galaxies in Sec.~V; that is, 
\begin{equation}\label{qr}
q(\bfr)=\frac{1}{4\pi\lambda}\,\frac{1}{|\bfr|^2}\,.
\end{equation}
It follows that
\begin{equation}\label{qk}
\hat{q}(\bfk)=\frac{1}{4(2\pi)^{1/2}\lambda}\,\frac{1}{|\bfk|}\,.
\end{equation}
Thus for any $M>0$, $M_{\rm D}=\infty$.

Once $q$ is determined from Eq. (\ref{convolution}), it is possible to
work out its reciprocal kernel. In fact, it follows from
Eqs.~(\ref{kernel3}) and (\ref{quu}) that
$\hat{p}_1(\bfk)=\hat{p}(\bfk)$,
\begin{equation}\label{peenplus1}
\hat{p}_{n+1}(\bfk)=-\nu\hat{p}(\bfk)\hat{p}_n(\bfk)\,,
\end{equation}
and
\begin{equation}\label{pee1}
-\hat{q}(\bfk)=\sum\limits_{n=1}^\infty\hat{p}_n(\bfk)\,.
\end{equation}
It is then straightforward to show that
\begin{equation}\label{quuhat}
{-\hat{q}(\bfk)=\frac{\hat{p}(\bfk)}{1+\nu\hat{p}(\bfk)}\,.}
\end{equation}
Therefore,
\begin{equation}\label{peehat}
{-\hat{p}(\bfk)=\frac{\hat{q}(\bfk)}{1+\nu\hat{q}(\bfk)}\,. }
\end{equation}
These results are consistent with the fact that $p$ and $q$ are
reciprocal of each other.


\begin{thebibliography}{99}

\bibitem{BahramAnnalen} B.~Mashhoon, 
Ann.\ Phys.\ (Berlin) {\bf 17}, 705 
(2008) [arXiv:0805.2926].

\bibitem{Bahram2007} B.~Mashhoon, 
  Ann.\ Phys.\ (Berlin) {\bf 16}, 57 
  (2007) [arXiv:hep-th/0608010].

\bibitem{Lochlainn} L.~O'Raifeartaigh, {\it The Dawning of Gauge
    Theory} (Princeton University Press, Princeton, NJ, 1997).

\bibitem{NitschErice} J.~Nitsch, 
  in {\it Proceedings of the 6th Course of the School of Cosmology and
    Gravitation on `Spin, Torsion, Rotation, and Supergravity',
  Erice, Italy, 1979,} edited by P.~G.~Bergmann and V.~de Sabbata
  (Plenum, New York, 1980) p.\ 63
  .

\bibitem{NitschHehl} J.~Nitsch and F.~W.~Hehl, 
  Phys.\ Lett. B {\bf 90}, 98 
  (1980).

\bibitem{Muench00} U.~Muench, F.~W.~Hehl, and B.~Mashhoon, 
  Phys.\ Lett.\ A {\bf 271}, 8 (2000) [arXiv:gr-qc/0003093].

\bibitem{Birkbook} F.~W.~Hehl and Yu.~N.~Obukhov, {\it Foundations of
    Classical Electrodynamics: Charge, Flux, and Metric}
  (Birkh\"auser, Boston, MA, 2003).

\bibitem{HehlHeld} F.~W.~Hehl, J.~Nitsch, and P.~Von der Heyde, in {\it
    General Relativity and Gravitation, Volume 1,} edited by A.~Held
  (Plenum, New York, 1980) p.\ 329
; see, in particular, the
  Appendix.

\bibitem{HehlErice79} F.~W.~Hehl, 
  in {\it Proc.\ of the 6th Course of the School of Cosmology and
    Gravitation on `Spin, Torsion, Rotation, and Supergravity',}
  Erice, Italy, 1979, edited by P.~G.~Bergmann and V.~de Sabbata
  (Plenum, New York, 1980) p.\ 5.

\bibitem{PRs} F.~W.~Hehl, J.~D.~McCrea, E.~W.~Mielke, and Y.~Ne'eman,
  Phys.\ Rep.\ {\bf 258}, 1 
(1995).

\bibitem{Schouten} J.~A.\ Schouten, {\it Tensor Analysis for
    Physicists,} 2nd ed.\ reprinted (Dover, Mineola, NY, 1989).

\bibitem{Aldrovandi1}
  R.~Aldrovandi, J.~G.~Pereira, and K.~H.~Vu,
  Gen.\ Relativ.\ Gravit.\  {\bf 36}, 101 (2004)
  [arXiv:gr-qc/0304106].

\bibitem{Itin1} Y.~Itin, 
Gen.\ Relativ.\ Gravit.\ {\bf 34}, 1819
  (2002) [arXiv:gr-qc/0111087].

\bibitem{Itin2} Y.~Itin, 
  in {\it Classical and Quantum Gravity Research Progress,} edited by
  M.~N.~Christiansen and T.~K.~Rasmussen (Nova Science Publishers,
  Hauppauge, NY, 2008) [arXiv:0711.4209].

\bibitem{Maluf1}
  J.~W.~Maluf, M.~V.~O.~Veiga, and J.~F.~da Rocha-Neto,
  Gen.\ Relativ.\ Gravit.\  {\bf 39}, 227 (2007)
  [arXiv:gr-qc/0507122].
 
\bibitem{Maluf2}
  J.~W.~Maluf and S.~C.~Ulhoa,
  Phys.\ Rev.\  D {\bf 78}, 047502 (2008)
  [arXiv:0807.0255].
 
\bibitem{Egg} E.~W.~Mielke, 
  Ann.\ Phys.\ (NY) {\bf 219}, 78 
(1992).

\bibitem{YuriTele1} Yu.~N.~Obukhov and J.~G.~Pereira, 
  Phys.\ Rev.\ D {\bf 67}, 044016 (2003) [arXiv:gr-qc/0212080].

\bibitem{YuriTele2} Y.~N.~Obukhov and G.~F.~Rubilar, 
  Phys.\ Rev.\ D {\bf 73}, 124017 (2006) [arXiv:gr-qc/0605045].

\bibitem{YuriTele3} Y.~N.~Obukhov, G.~F.~Rubilar, and J.~G.~Pereira,
  Phys.\ Rev.\ D {\bf 74}, 104007 (2006) [arXiv:gr-qc/0610092].

\bibitem{So} L.~L.~So and J.~M.~Nester, 
  in {\it Proceedings of the 10th Marcel Grossmann Meeting, Rio de
    Janeiro, Brazil, 2003}, edited by M.~Novello et
  al. (World Scientific, Hackensack, NJ, 2005), Part B, p.\ 1498
  [arXiv:gr-qc/0612062].

\bibitem{Tung} R.~S.~Tung and J.~M.~Nester, 
  Phys.\ Rev.\ D {\bf 60}, 021501 (1999) [arXiv:gr-qc/9809030].

\bibitem{Milutin} M.~Blagojevi\'c, {\it Gravitation and Gauge
    Symmetries} (IoP, Bristol, UK, 2002).

\bibitem{Ortin} T.~Ort{\'\i}n, {\it Gravity and Strings}
  (Cambridge University Press, Cambridge, UK, 2004).

\bibitem{Soussa2003}
  M.~E.~Soussa and R.~P.~Woodard,
  Class.\ Quant.\ Grav.\ {\bf 20}, 2737 
  (2003) [arXiv:astro-ph/0302030].

\bibitem{Barvinsky2003}
  A.~O.~Barvinsky,
  Phys.\ Lett.\  B {\bf 572}, 109 
(2003)
  [arXiv:hep-th/0304229].

\bibitem{Synge} J.~L.~Synge, {\it Relativity: The General Theory}
  (North-Holland, Amsterdam, 1971).

\bibitem{Edelen} D.~G.~B.~Edelen, {\em Nonlocal Variations {\em and}
    Local Invariance of Fields} (American Elsevier, New York, 1969).

\bibitem{Minkowski} H.~Minkowski, 
  K\"onigliche Gesellschaft der Wissen\-schaf\-ten zu G\"ottingen.
  Mathematisch-Physikalische Klasse. Nachrichten, p.~53 (1908).

\bibitem{Chicone07}
  C.~Chicone and B.~Mashhoon,
  Ann.\ Phys.\ (Berlin) {\bf 16}, 811 (2007)
  [arXiv:0708.2744].

\bibitem{Rubin} V.~C.~Rubin and W.K. Ford, Astrophys.\ J. {\bf 159},
  379 (1970).  

\bibitem{Roberts} M.~S.~Roberts and R.~N.~Whitehurst, Astrophys.\
  J. {\bf 201}, 327 (1975).

\bibitem{Sofue} Y.~Sofue and V.~Rubin, 
Annu.\ Rev.\ Astron.\ Astrophys. {\bf 39}, 137 
    (2001).

\bibitem{Carignan} C.~Carignan, L.~Chemin, W.~K.~Huchtmeier, and
  F.~J.~Lockman, 
  Astrophys.\ J.\ Lett. {\bf 641}, L109 
  (2006); P.~Salucci et al., Mon.\ Not.\ R.\ Astron.\ Soc.\ {\bf 378},
  41 (2007).

\bibitem{Jacob2004} J.~D.~Bekenstein, Phys.\ Rev.\ D {\bf 70}, 083509
  (2004); erratum, Phys.\ Rev.\ D {\bf 71}, 069901(E) (2005)
  [arXiv:astro-ph/0403694].
 
\bibitem{Jacob2006} J.~D.~Bekenstein, 
 Contemp.\ Phys. {\bf 47}, 387 
  (2006) [arXiv:astro-ph/0701848].

\bibitem{Bruneton2008} J.~P.~Bruneton, S.~Liberati, L.~Sindoni, and
  B.~Famaey,
  [arXiv:0811.3143].

\bibitem{TohlineKuhn} J.~E.~Tohline, in {\it IAU Symposium 100,
    Internal Kinematics and Dynamics of Galaxies,} edited by
  E.~Athanassoula (Reidel, Dordrecht, 1983), p.~205. J.~R.~Kuhn and
  L.~Kruglyak, 
  Astrophys.\ J. {\bf 313}, 1 
  (1987).

\bibitem{Jacob1988} J.~D.~Bekenstein, in {\it Second Canadian
    Conference on General Relativity and Relativistic Astrophysics},
  A.~Coley, C.~Dyer, and T.~Tupper, eds. (World Scientific, Singapore,
  1988), p.~68.

\bibitem{Tricomi} F.~G.~Tricomi, {\it Integral Equations}
  (Interscience, New York, 1957).

\end{thebibliography}
\end{document}